\magnification=1200
\vsize=7.5in
\hsize=5in
\tolerance 10000
\pageno=1

\def\half{{1\over 2}}
\def\R{{\bf R}}
\def\A{{\bf A}}
\def\P{{\bf P}}
\def\sig{{\bf \sigma}}
\def\PRi{{\vert \Psi_i (\R )\rangle}}
\def\proji{{{\bf \Pi}_i}}
\def\proj+{{{\bf \Pi}_+}}
\def\proj-{{{\bf \Pi}_-}}
\def\phii{{\phi_i}}
\def\del{{{\bf \nabla}}}
\def\proj{{{\bf \Pi}}}
\baselineskip 12pt plus 2pt minus 2pt \centerline{\bf \quad
BERRY'S PHASE}
\bigskip
Berry's phase [1] is a quantum phase effect arising in systems that undergo a slow,
cyclic evolution.  It is a remarkable correction to the quantum adiabatic theorem and
to the closely related Born-Oppenheimer approximation [2].  Berry's elegant and general
analysis has found application to such diverse fields as atomic, condensed matter,
nuclear and elementary particle physics, and optics. In this brief review, we first
derive Berry's phase in the context of the quantum adiabatic theorem and then in the
context of the Born-Oppenheimer approximation.  We mention generalizations of Berry's
phase and analyze its relation to the $\rightarrow$Aharonov-Bohm effect.

Consider a Hamiltonian $H_f(\R )$ that depends on parameters $R_1, R_2, \dots, R_N$,
components of a vector $\R$.  Let us assume that $H_f (\R )$ has at least one discrete
and nondegenerate eigenvalue $E_i (\R )$ with $\vert \Psi_i (\R )\rangle$ its
eigenstate; $E_i (\R )$ and $\vert \Psi_i (\R ) \rangle$ inherit their dependence on
$\R$ from $H(\R )$.  If the vector $\R$ changes in time, then $\vert \Psi_i (\R
(t))\rangle$ is not an exact solution to the time-dependent Schr\"odinger equation. But
if $\R$ changes slowly enough, the system will not $\rightarrow$jump to another
eigenstate. Instead, it adjusts itself to the changing Hamiltonian.  A heavy weight
hanging on a string illustrates such adiabaticity.  Pull the string quickly---it snaps
and the weight falls. Pull the string slowly---the weight comes up with it.

``Slowly enough" has the following formal sense.  Let $\R [t/T]$ evolve over a time
interval $0\le t \le T$; the larger $T$, the slower the evolution.  If at time $t=0$
the system is in the state $\vert \Psi_i (\R [0] ) \rangle$, then at time $t=T$ the
state is $e^{i\phii (T)} \vert \Psi_i (\R [1])\rangle$ with probability approaching 1
as $T$ approaches infinity, according to the quantum adiabatic theorem [10].  We obtain
the phase $\phii (t)$ by substituting $e^{i\phii (t)} \vert \Psi_i (\R )\rangle$ into
the time-dependent $\rightarrow$Schr\"odinger equation,
$$
i\hbar {d \over { dt}} e^{i\phii (t)} \vert \Psi_i (\R )\rangle = H_f (\R [t/T])
e^{i\phii (t)} \vert \Psi_i (\R)\rangle~~~,
$$
and projecting both sides of the equation onto $e^{i\phii (t)} \vert \Psi_i (\R
)\rangle$:
$$
{{d}\over {dt}} \phii (t) = {i}\langle \Psi_i (\R ) \vert
\del_\R \vert \Psi_i (\R ) \rangle \cdot {{d\R} \over {dt}}
-{1\over\hbar} E_i (\R ) ~~~~.
$$
Thus
$$\eqalign{
\phii (t) -\phii (0) &=\int_0^t dt^\prime \left[ i\langle \Psi_i
(\R ) \vert \del_\R \vert \Psi_i (\R ) \rangle \cdot {{d\R}\over
{dt^\prime }}- {1\over\hbar} E_i (\R ) \right]\cr &= \int_{\R
[0]}^{\R [t]} \langle \Psi_i (\R) \vert i\del_\R \vert \Psi_i (\R
) \rangle \cdot d\R -{1\over \hbar} \int_0^t dt^\prime ~E_i (\R
)~~~~.\cr }
$$
The integrand $\A_B \equiv \langle \Psi_i (\R )\vert i\del_\R \vert \Psi_i(\R )\rangle$
is {\it Berry's connection} for the state $\vert \Psi_i (\R )\rangle$. The integral
$-\int_0^t E_i dt^\prime / \hbar$ is called the {\it dynamical} phase.

The overall phase of a quantum state is not observable.  But a quantum system may be in
a $\rightarrow$superposition of states; the {\it relative} phase of these states is
observable. Consider two paths $\R [t/T]$ and $\R^\prime [t/T]$ with the same endpoints
$\R [0] =\R^\prime [0]$ and $\R [1]= \R^\prime [1]$, and suppose that the system
evolves in a superposition of states $\vert \Psi_i (\R [t/T])\rangle$ and $\vert
\Psi_i(\R^\prime [t/T]) \rangle$. At time $t=T$ the relative phase of this
superposition contains two parts. One part is the relative dynamical phase. The other
part is Berry's phase, the difference between $\A_B$ integrated along $\R$ and $\A_B$
integrated along $\R^\prime$, i.e. it is the circular integral of $\A_B$ along the {\it
closed} path comprising $\R$ and $\R^\prime$ with opposite senses.  This phase is well
defined, because it is $\rightarrow$gauge invariant:  If we multiply $\vert \Psi_i (\R
)\rangle$ by a phase factor $e^{i\Lambda (\R )}$, it remains the same instantaneous
eigenstate of $H_f (\R )$, but $\A_B$ changes by $-\del_\R \Lambda (\R )$. Since the
change in $\A_B$ is a gradient, the integral of $\A_B$ around a closed loop is
unchanged, hence well defined.

As an example of Berry's phase, consider the spin-1/2 Hamiltonian \hbox{$H_f=\mu \R
\cdot \sig$}, where $\sigma_x$, $\sigma_y$ and $\sigma_z$ are the $\rightarrow$Pauli
spin matrices. The eigenstate corresponding to the positive eigenvalue $E_+ = \mu R$ is
$$
\pmatrix{  \cos {\theta \over 2}\cr
 e^{i\phi} \sin {\theta \over 2}\cr }
~~~,
$$
where $R_z =R\cos \theta$ and $R_x +iR_y = Re^{i\phi}\sin \theta$. The Berry
connection, expressed as a function of $\theta$ and $\phi$, is $(\A_B)_\theta =0$,
$(\A_B)_\phi = (\cos \theta -1)/2$ and matches the vector potential of a Dirac monopole
of strength 1/2 located at the origin $\R={\bf 0}$.  The integral of $\A_B$ along any
loop in $\R$ equals $-1/2$ times the solid angle subtended by the loop at the origin
(as an application of Stokes's theorem shows).  This example is generic because wherever
two nondegenerate energy levels cross at a point in a space of parameters, the
Hamiltonian near the point reduces to an effective two-level Hamiltonian proportional
to $\R \cdot \sig$, with the degeneracy at $\R={\bf 0}$. Hence an effective magnetic
monopole can arise wherever two discrete, nondegenerate levels become degenerate.

The spin-1/2 example also illustrates how Berry's phase can be topological.  A loop in
$\R$ defines two solid angles, just as a loop on the surface of a sphere cuts the
surface into two parts.  Why, then, is Berry's phase not ambiguous?  The answer is that
the difference between the two solid angles is equal to $\pm 4\pi$.  (The two solid
angles have opposite signs because their orientations, or the directions of integration
of $\A_B$, are opposite.) But a $\pm 4\pi$ difference of solid angle corresponds to a
$\mp 2\pi$ difference in phase, which is unobservable.  Here Berry's phase obeys a
constraint arising from the topology of a sphere.

In the Born-Oppenheimer approximation, the $R_1, R_2, \dots$ are quantum observables
and may not even commute. They evolve according to their own ``slow" Hamiltonian $H_s$,
and the overall Hamiltonian is the sum $H=H_f+ H_s$.  The eigenvalues of $H_f$ must be
discrete, and the adiabatic limit applies if $H_s$ is an arbitrarily weak perturbation
on $H_f$.  The weaker the perturbation, the smaller the probability of transitions
(quantum jumps) among the eigenstates of $H_f$.  The unperturbed Hilbert space for $H$
divides into subspaces, one for each eigenvalue $E_i$ of $H_f$. In the adiabatic limit,
the ``fast" variables remain in an eigenstate $\PRi$ of $H_f$, with $i$ fixed, while
dynamical and Berry phases of $\PRi$ show up in $H$ as induced scalar and vector
potentials.

Born and Oppenheimer multiplied $\vert \Psi_i (\R )\rangle$ by a function $\Phi (\R
,t)$ and obtained an effective Hamiltonian for $\Phi (\R ,t)$.  Here we obtain and
simplify their effective Hamiltonian algebraically.  Let $\proji$ denote the
$\rightarrow$operator of projection onto the subspace corresponding to $E_i$. The
subspaces are disjoint and form a complete set: $\sum_i \proji =1$.  In the adiabatic
limit, we can replace $H_s$ by $\sum_i \proji H_s \proji$ to obtain the effective
Hamiltonian of Born and Oppenheimer:
$$
H_{eff} = H_f + \sum_i \proji H_s \proji  ~~~~.
$$
In $H_{eff}$ there are induced potentials.  If
$$
H_s ={{P^2} / {2M}} + V(\R )~~~,
$$
where $P_i =-i\hbar \partial /\partial R_i$, the sum $\sum_i \proji H_s \proji$ in
$H_{eff}$ contains products of the form
$$
\proji P^2 \proji = \sum_{j} \proji \P \proj_j \P \proji~~~~.
$$
We simplify them by decomposing $\P$ into two parts, $\P = (\P -\A )+ \A$.  The first
part acts only {\it within} subspaces; that is, $[\P - \A , \proji ] = 0$ for all $i$.
Only the second part, $\A$, causes transitions among the subspaces. Like a vector
potential, $\A$ is somewhat arbitrary: we can add to $\A$ any term that commutes with
the $\proji$. Let us remove this arbitrariness by requiring $\proji \A \proji = 0$ for
each $i$.  The effective Hamiltonian for the $\R$ is then [3]
$$
H_{eff} = H_f + {1 \over {2M}} (\P - \A )^2 +  {1 \over {2M} }
\sum_i \proji \A^2 \proji +V(\R )~~~~.
$$
The sum in $i$ is an induced scalar potential, while $\A$ is an induced vector
potential:  $\A$ is Berry's connection $\A_B$ in an off-diagonal gauge.  For example,
suppose $H_f=\mu \R \cdot \sig$ as in the spin-1/2 example above.  The operators of
projection corresponding to $E_\pm =\pm \mu R$ are
$$
{\bf \Pi}_\pm = \half (1\pm \R \cdot \sig /R)~~~,
$$
and the vector potential
$$
\A = {{\hbar \R\times \sigma}\over{2R^2}}
$$
solves the two conditions $[\P -\A ,{\bf \Pi}_\pm ] =0$ and ${\bf \Pi}_\pm\A {\bf
\Pi}_\pm=0$; $\A$ is off-diagonal.  The field corresponding to $\A$,
$$
B_i = {1 \over 2} \epsilon_{ijk} F_{jk} ={1\over 2} \epsilon_{ijk} (\partial_j A_k -
\partial _k A_j - i [ A_j , A_k ] ) = - { \hbar {R_i} \over {2R^4}}(\R \cdot \sig )
~~~,$$ is a monopole field ${\bf B}=\mp \hbar \R /2R^3$ since the eigenvalues of $\R
\cdot\sig/R$ are $\pm 1$.
\goodbreak

So far we have taken the eigenvalues of $H_f$ to be discrete and nondegenerate.  If
$H_f$ has a discrete and {\it degenerate} eigenvalue, Berry's phase may be non-abelian
[4].  The eigenstates belonging to this eigenvalue do not (in the adiabatic
approximation) jump to eigenstates belonging to other eigenvalues, but they may mix
among themselves.  The mixing amounts to multiplication by a non-abelian phase, i.e. a
unitary matrix.

Another generalization of Berry's phase is the Aharonov-Anandan phase [5].  Suppose a
system evolves according to Schr\"odinger's equation, but the change in the Hamiltonian
is neither adiabatic nor cyclic.  Aharonov and Ananden showed that the system can still
exhibit a Berry phase; all that is needed is cyclic evolution of the {\it state} of the
system. Cyclic evolution of a state defines a closed path in the Hilbert space of the
state. Whether or not this evolution is adiabatic, it leaves the system with a
dynamical phase, which depends on the Hamiltonian of the system, and a geometrical
phase---Berry's phase---which depends only on the closed path of the state in its
Hilbert space.  Thus Berry's phase need not be adiabatic (although it is still a
correction to the adiabatic theorem).

We have considered evolution consistent with Schr\"odinger's equation.  But as
Pancharatnam showed [6], geometric phases can emerge from nonunitary evolution. For
example, let an ensemble be divided into two subensembles, one of which is subjected to
a sequence of filtering measurements (projections).  If the sub-subensemble that
survives this filtering has returned to its initial state, it has a well defined phase
(relative to the unfiltered subensemble) which equals a relative dynamical phase plus
the Berry phase for this evolution.

Berry's phase has a classical analogue:  {\it Hannay's angle} [7] is a phase effect in
a classical periodic system that depends on adiabatically changing parameters.  A
canonical pair of variables for such a system is an ``action" variable $I$, which is an
adiabatic constant of the motion, and a conjugate ``angle" variable $\phi$.  Hannay's
angle is an extra shift in $\phi$ acquired by the system during a cyclic evolution in
the space of parameters.  When the Hannay angle of a system depends on its action $I$,
the corresponding quantum system acquires a Berry phase during the same cyclic
evolution [8].

Although the Aharonov-Bohm effect has no classical analogue, we may treat it as an
example of Berry's phase.  More generally, however, the Aharonov-Bohm and Berry phases
can {\it combine} in a topological phase [9].  For example, imagine a ``semifluxon",
something like a straight, heavy, infinite solenoid enclosing flux $hc/2e$---exactly
half a flux quantum---that moves perpendicular to itself. It interacts with an electron
wave function that has support in two disjoint regions. If the semifluxon moves in a
slow circuit, we can ask what phase the electron acquires from this adiabatic cyclic
evolution. Figure 1 shows one of the two regions where the electron wave function has
support, and two possible circuits for the semifluxon. If the semifluxon evolves along
$C_1$, the electron acquires no relative Berry phase and also the Aharonov-Bohm phase
vanishes. If the semifluxon evolves along $C_2$, the relative Berry phase is $\pi$ and
it is entirely the Aharonov-Bohm phase. If the semifluxon does neither but plows
through the electron wave function, we might expect the Berry phase to lie between 0
and $\pi$.  However, it can be shown (using time-reversal symmetry) that the Berry
phase can only be $0$ or $\pi$.  Since the path of the semifluxon is arbitrary, at some
point ${\cal P}$, the Berry phase must jump from 0 to $\pi$, i.e. the electron wave
function must become degenerate when the semifluxon is situated at ${\cal P}$. Here the
Berry phase and the Aharonov-Bohm phase combine in a single topological phase that
depends only on the winding number of the semifluxon path around the point ${\cal P}$.
\bigskip

\centerline{\bf \quad References}
\medskip
\noindent [Primary]

[1] {M. V. Berry, ``Quantal phase factors accompanying adiabatic changes", {\it Proc.
R. Soc. Lond.}, {\bf A392}, 45-57 (1984).}

[2] {M. Born and J. R. Oppenheimer, ``Zur Quantentheorie der Molekeln", {\it Annalen
der Physik} {\bf 84}, 457-484 (1927); see also C. A. Mead and D. G. Truhlar, ``On the
determination of Born-Oppenheimer nuclear motion wave functions including complications
due to conical intersections and identical nuclei", {\it J. Chem. Phys.} {\bf 70},
2284-2296 (1979).}

[3] Y. Aharonov, E. Ben-Reuven, S. Popescu and D. Rohrlich, ``Perturbative induction of
vector potentials", {\it Phys. Rev. Lett.} {\bf 65}, 3065-3067 (1990) and {\bf 65}, 863
(1992); ``Born-Oppenheimer revisited", {\it Nucl. Phys.} {\bf B350}, 818-830 (1991).

[4] {F. Wilczek and A. Zee, ``Appearance of gauge structure in simple dynamical
systems",  {\it Phys. Rev. Lett.} {\bf 52}, 2111-2114 (1984); A. Zee, ``Non-Abelian
gauge structure in nuclear quadrupole resonance", {\it Phys. Rev.} {\bf A38}, 1-6
(1988).}

[5] {Y. Aharonov and J. Anandan, ``Phase change during a cyclic quantum evolution",
{\it Phys. Rev. Lett.} {\bf 58}, 1593-1596 (1987).}

[6] S. Pancharatnam, ``Generalized theory of interference, and its applications. Part
I. Coherent pencils ", {\it Proc. Ind. Acad. Sci.} {\bf A44}, 247-262 (1956).

[7] {J. H. Hannay, ``Angle variable holonomy in adiabatic excursion of an integrable
Hamiltonian", {\it J. Phys. A: Math. Gen.} {\bf 18}, 221-230 (1985).}

[8] {M. V. Berry, ``Classical adiabatic angles and quantal adiabatic phase", {\it
\hbox{J. Phys. A:} Math. Gen.} {\bf 18}, 15-27 (1985).}

[9] {Y. Aharonov, S. Coleman, A. Goldhaber, S. Nussinov, S. Popescu, B. Reznik, D.
Rohrlich, and L. Vaidman, ``Aharonov-Bohm and Berry phases for a quantum cloud of
charge", {\it Phys. Rev. Lett.} {\bf 73}, 918-921 (1994).}

\medskip
\noindent [Secondary]

[10] {See A. Messiah, {\it Quantum Mechanics}, Vol. {\bf II}, trans. J. Potter
(North-Holland, Amsterdam, 1963), Chap. XVII, Sects. 10-12.}

\bigskip

\centerline{\bf \quad Figure Caption}
\medskip
Fig. 1.  An electron cloud with support in a region ${\cal S}$ (and in disjoint region
not shown) and two possible paths, $C_1$ and $C_2$, of a semifluxon.  At the point
${\cal P}$, the semifluxon induces a degeneracy in the energy of the electron.

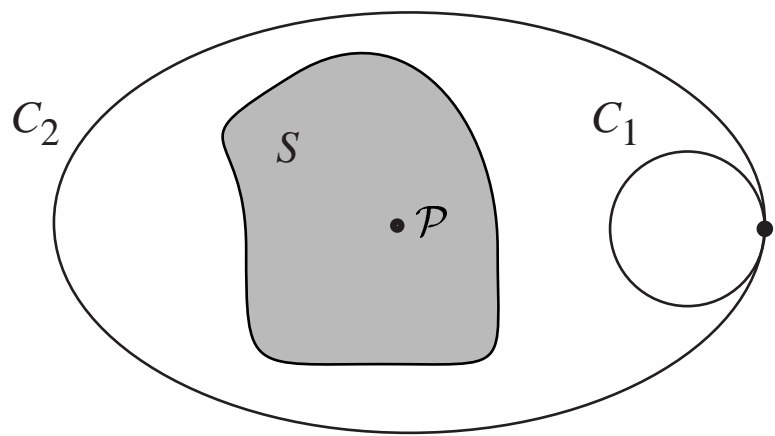
\end